\documentclass[preprint,12pt,authoryear]{elsarticle}
\pdfoutput=1

\usepackage{natbib}
\usepackage{amssymb}
\usepackage{float}
\usepackage[version=4]{mhchem}
\usepackage{multirow}

\begin{document}
\begin{frontmatter}



\title{Expanding swirl combustor operability on methane-ammonia-air mixtures using a distributed fuel injection technique and inlet air preheating}



\author{V. Viswamithra, M. Gurunadhan, S. Menon}

\affiliation{organization={Louisiana State University, Mechanical Engineering Department},
            addressline={3261 Patrick F Taylor Hall}, 
            city={Baton Rouge},
            postcode={70803}, 
            state={LA},
            country={USA}}

\begin{abstract}
The necessity to reduce greenhouse gas emissions has prompted the search for carbon-free fuel alternatives. One such carbon-free fuel source that can be used for power generation in a gas turbine-based system is ammonia. Ammonia combustion poses challenges due to reduced flame speed and reactivity, which can be alleviated by addition of methane or natural gas. However, \ce{NOx} emissions continue to be an issue, which needs resolution before practical consideration of ammonia as a fuel source. One of the potential solutions that has been proposed is a two-stage, rich-lean combustion process to minimize \ce{NOx} while ensuring complete reactant consumption. This work evaluates the use of two strategies to widen stability limits for swirl combustors operating on premixed methane-ammonia-air mixtures, which would facilitate the two-stage combustion approach. The first strategy involves the use of a distributed fuel injection approach utilizing a novel micro fuel injection swirler to facilitate homogeneous mixing in a highly compact manner while preventing flashback concerns. The second strategy involves use of inlet air preheating to increase flame stability and delay blow-off. Experiments and reactor network simulations are carried out to evaluate the effectiveness of the two strategies in expanding combustor operability limits. The corresponding influence of the proposed strategies on \ce{NOx} emissions are studied and underlying reaction pathways leading to \ce{NOx} production are analyzed. Results of the study indicate that a distributed fuel injection strategy is able to significantly expand stability limits of a swirl combustor operating on methane-ammonia-air mixtures. Inlet air preheating provides additional expansion of stability limits, however, this is accompanied by increased \ce{NOx} production, which is undesirable from an emissions standpoint. \ce{NOx} is found to be significantly lower for rich fuel-air mixtures primarily through the effect of NHi reaction pathways responsible for \ce{NO} consumption facilitated by the presence of excess \ce{NH3}. 


\end{abstract}




\end{frontmatter}


\section{Introduction}
Ammonia (\ce{NH3}) is a carbon-free molecule that has the potential to provide an alternative fuel pathway to reduce green house gas (GHG) emissions generated by hydrocarbon fuels~\cite{kobayashi2019science}. Renewable energy sources can be used to generate Ammonia avoiding the energy intense Haber Bosch process, thus ensuring a lower life-cycle GHG emission~\cite{smith2020current}. However, direct use of ammonia for power generation in a gas turbine combustor is challenged by low flame speed and issues with flame stability~\cite{verkamp1967ammonia}. These issues limit the lean blow-off limit (LBO) for ammonia-air flames which is undesirable from a soot and combustion efficiency standpoint. Efforts have been made to improve the combustion characteristics of ammonia by adding methane (\ce{CH4}) to the mixture. While the addition of methane has resulted in improved flame stability and reduced LBO, Nitrous Oxide (\ce{NOx}) emissions have been found to be significantly higher than those from conventional fuels such as natural gas~\cite{khateeb2020stability,zhang2021emission,valera2017ammonia}. A recent study by this group considered the use of a novel Micro Fuel Injection Swirler (MFIS) module to facilitate improvements in ammonia-methane-air mixtures in a swirl combustor~\cite{viswamithra2022distributed}. The particular advantage of this swirler is its ability to achieve high quality mixing of fuel with air in a very short distance ($\sim 2.5$ cm). Enhanced mixing is achieved using a distributed fuel injection strategy, where a large number (1056) of small fuel injection holes ($D=150~\mu$m) are machined directly into the swirler blades~\cite{giglio2009distributed}. 
This swirler strategy facilitates two key advantages. One, fuel and air are mixed just upstream of the flame zone, and with the swirler itself being water-cooled, flashback concerns are significantly reduced. Second, more uniform mixing is achieved through distributed injection, which is further augmented by turbulence-aided mixing by air crossflow over the mixing holes. These advantages facilitated a significant expansion of the combustor operability regime, allowing operation at lean mixtures with equivalence ratios ($\phi$) lower than those currently achievable with conventional swirlers, while allowing for high ammonia content (80-90\%). 

However, \ce{NOx} levels obtained in the previous study (1500--3500 ppm for power levels from 6--20 kW) were still significantly higher than the $<50$ ppm levels targeted by turbine engine manufacturers~\cite{tanaka2013development}. Recent studies have pursued two-stage, rich-lean combustion approaches to mitigate \ce{NOx} emissions~\cite{kurata2019development,pugh2019influence,okafor2019measurement,okafor2020control}. 
These studies emphasize the importance of efficient fuel-air mixing in the rich primary zone to ensure complete oxidation of ammonia~\cite{okafor2019towards,somarathne2017numerical}. 
Studies have reported a range of optimal equivalence ratios for the primary zone from 1.1~\cite{hayakawa2017experimental}, 1.2~\cite{somarathne2017numerical}, 1.5~\cite{li2019analysis}, to 1.3-1.35~\cite{okafor2020control}. This could be due to differences in combustor geometry and fuel injection strategies. However, the variation in optimal primary $\phi$ illustrates that there is a need to ensure operation across a sufficiently wide band of $\phi$ in the primary zone. Further, there is a need to ensure homogeneous mixing in the rich primary zone, including close to combustor walls~\cite{okafor2020control} to suppress \ce{NOx} generation in the secondary.
The present study aims to demonstrate the ability of the novel MFIS swirler to extend operability regime at rich and lean equivalence ratios through its ability to improve mixing quality. Operational range at rich $\phi$, particularly at low \ce{NH3} content which is limited by flashback issues will be attempted to be expanded. 
Besides extending the rich and lean limits, it is also desired to allow for combustion of high ammonia content mixtures close to 100\%. The capabilities of the MFIS swirler to expand operational range primarily through improved mixing will be supplemented by studies considering increase in inlet air temperature by pre-heating. Pre-heating of reactant mixtures has shown to be a viable strategy to suppress LBO as well as reduce unburnt ammonia content at rich mixtures~\cite{li2016research}. However, use of this strategy needs careful consideration for its effect on \ce{NOx} formation.

The layout of the paper is as follows. The experimental setup used in the investigation of methane-ammonia-air flames with the MFIS injection strategy in a model swirl combustor with air preheating is described along with the diagnostics employed and test conditions. The simulation approach based on reactor network modeling using flowfield information generated by a more detailed fluid flow simulation of the swirl combustor is discussed. Results obtained from experiments and simulations for blow-off limits and \ce{NOx} generation are presented and discussed. Sensitivity of \ce{NOx} formation reaction pathways to $\phi$, ammonia addition, and inlet air temperature is analyzed using reactor network simulations. Finally, conclusions are presented along with future research directions.

\section{Experimental Approach}
Experiments were conducted on a premixed swirl combustor setup operating on gaseous fuels. Combustor rig components, measurement diagnostics, test case matrix, and experimental approach are discussed next.

\subsection{Micro Fuel Injection Swirler (MFIS)} 
\begin{figure}[h!]
	\centering
	\includegraphics[scale=0.55]{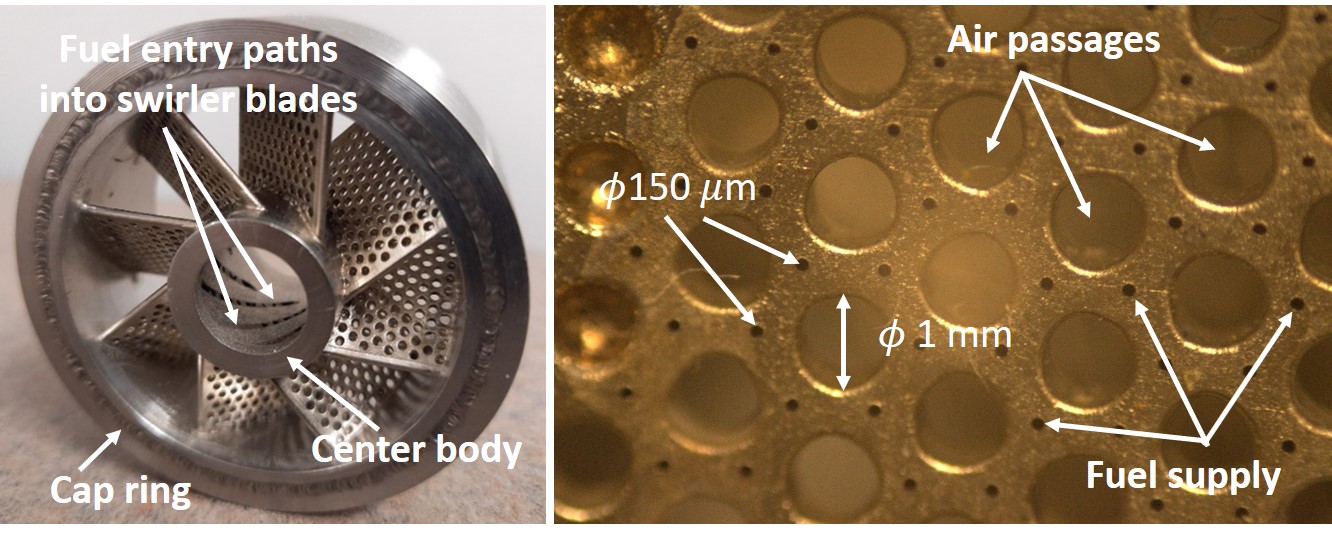}
	\caption{Micro fuel injection swirler (MFIS).}
	\label{fig:SwirlerPictures}
\end{figure}
The MFIS facilitates improved fuel-air mixing in a very short mixing length, while providing protection from flashback occurrence in the model combustor. The swirler shown in Fig.~\ref{fig:SwirlerPictures}, with a geometric swirl number of 0.75, incorporates a large number of fuel injection points directly located on the swirler vanes. Air entering the swirler flows over the vanes as well as within through holes located on the vanes. A fuel injection tube located within the central channel supplies fuel to the injection points on the swirler vanes through channels machined within the vanes. This architecture results in the generation of a large number of injection points, more than 1000, where fuel is injected into the cross-flowing air aided by the air flow occurring via through holes located on the swirler blade. As shown in the close-up photographs of the swirler blades in Fig.~\ref{fig:SwirlerPictures}, each fuel injection hole is surrounded by several holes carrying air flowing through the cross-section of the blade. The shear flow generated by this flow configuration along with the injection of the shear layers into a crossflow of air flowing over the vanes results in the generation of fine-scale turbulence aiding the fuel-air mixing process. The result is a more homogeneously mixed fuel-air mixture which can provide protection from flashback as well as improved resistance to blow-off by increasing flame stability. Manufacturing details of the MFIS have been reported by Giglio~\cite{giglio,giglio2009distributed}. 
\subsection{Model combustor rig}
\begin{figure}[h!]
	\centering
	\includegraphics[scale=0.5]{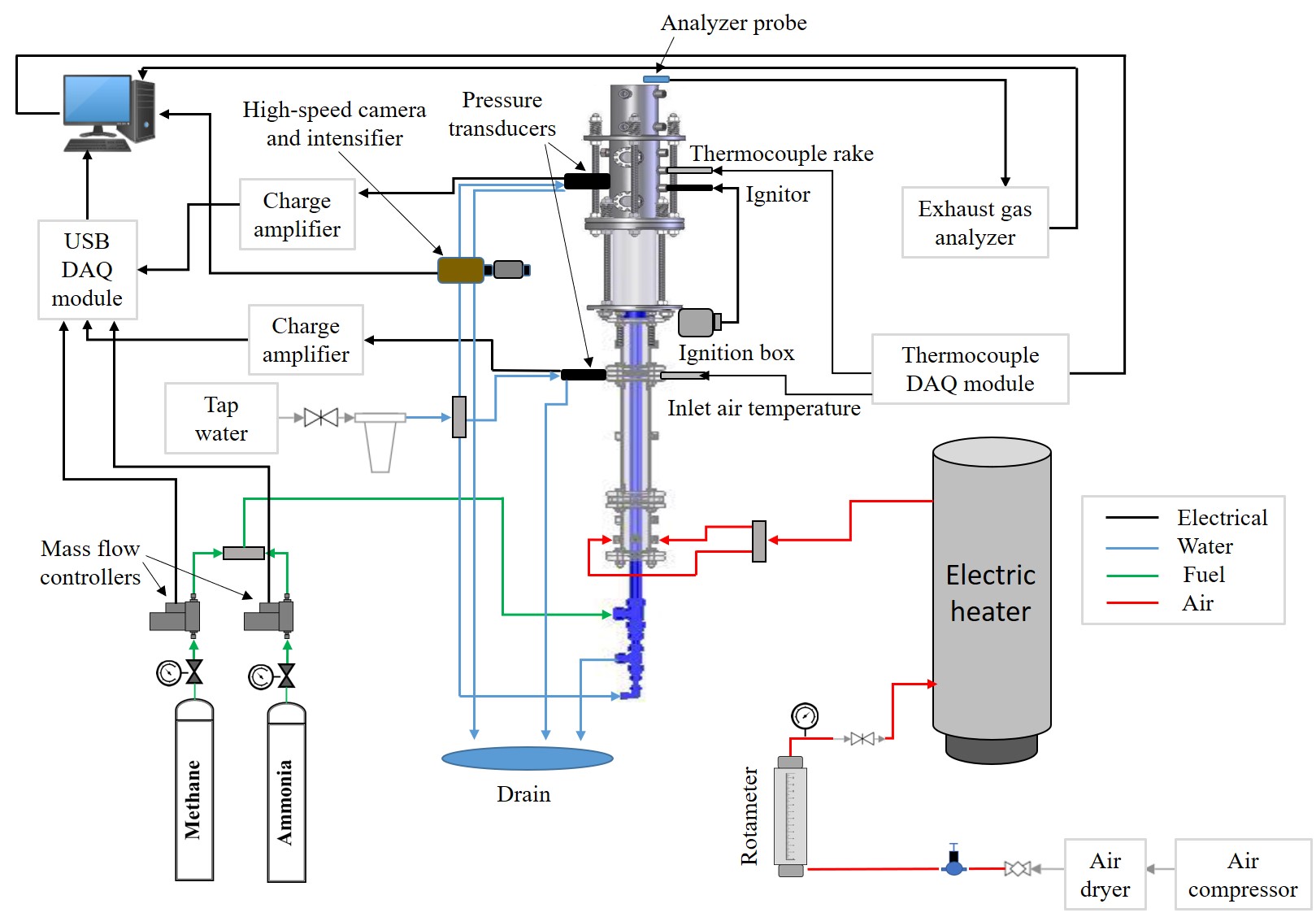}
	\caption{Model swirl combustor test rig with inlet air preheating.}
	\label{fig:TstRg}
\end{figure}
A model combustor setup is used to conduct tests with methane-ammonia-air mixtures using the MFIS swirler as discussed in the previous sub-section. Figure~\ref{fig:TstRg} shows a schematic of the experimental setup. Air from an external accumulation tank (2560 gal) provides air to the combustor rig. Air passes through a dryer, a globe valve, a pressure regulator (0-160psi) and then through a rotameter (King instrument Co.). The rotameter has a range of 5-40 SCFM (0.002-0.02 $m^3/s$). A pressure gauge immediately downstream of the rotameter is used to measure back-pressure and correct the air flow rate. An industrial air heater (Chromalox Model GCHB-2485) heats the air to the desired temperature. The coil based air heater is capable of heating air to temperatures up to 750$^0$C. A 2" diameter pipe carrying hot air to the combustor is insulated using fiberglass insulation. Heated air enters the upstream section of the model combustor comprising of concentric pipes which deliver fuel and air to the combustion chamber. Air supply is connected to the combustor's inlet ports through flexible metallic hoses. Before reaching the swirler, a flow straightener ensures that airflow is parallel to the pipe. A type-K thermocouple is installed upstream of the combustion chamber to measure inlet air temperature. Methane (Ultra-high purity, 99.99\%) and ammonia (Ultra-high purity, 99.99\%) are supplied from compressed gas cylinders using Omega FMA-5400 mass flow controllers. For ammonia, a flow controller with a range of 0–100 LPM (0–0.0016 $m^3/s$) is used, while for methane, a flow controller with a range of 0–50 LPM (0–0.008 $m^3/s$) is utilized. Both flow controllers are nitrogen-calibrated and have a full scale accuracy of $\pm$ 1.5\%. An upstream pressure of 138 kPa (20 psi) is set for both flow controllers. The mass flow controllers are controlled through a LabVIEW program and used to set the fuel-air mixture ratio. Fuel mixture enters the model combustor at the upstream section through a concentric pipe. Fuel and air move through the upstream section and enter the swirler region, where the MFIS as shown in Fig.~\ref{fig:SwirlerPictures} is used to generate a fuel-air mixture exiting 0.5 inches upstream of the dump plane of the combustor. The swirler is cooled by cooling water supplied through channels located just outside the fuel channel. A high voltage ignition source provides the ability to ignite the fuel-air mixture. Optical access is provided by the combustor, which is used to visualize combustion and heat release processes~\cite{viswamithra2022distributed}. This is accomplished using a 22.8 cm (9") long fused quartz tube with an inner diameter of 8.26 cm (3.25") and a thickness of 0.64 cm (0.25"). A thermocouple rake consisting of 4 thermocouples held about 30 cm above the dump plane is used to measure exhaust gas temperature. The 4 thermocouples are held at different radial distances from the axis of combustor. 

\subsection{Emissions Analyzer} The model combustor exhaust is sampled using an Enerac model 700 emissions analyzer. The analyzer which utilizes non-dispersive infrared (NDIR) technology can measure emissions of 5 gases: \ce{CO}, \ce{NO}, \ce{NO2}, \ce{SO2}, and \ce{O2} simultaneously. A probe with a built-in temperature sensor is connected to the exhaust from the combustor about 60 cm downstream of the dump plane. Measurements from the analyzer are sent to a computer where it is logged by the analyzer software at a rate of 60 Hz. The analyzer has an accuracy of ±2\%.
\subsection{Data acquisition and control}
 Gas mass ﬂow controllers operate on a 0-5 VDC analog voltage signal generated using a National Instruments (NI) USB-6343 multi-function data acquisition and control device (DAQ). A LabVIEW program running at 1 Hz is used to cycle through the test cases and send required output voltage to the mass ﬂow controller. Air flow rate is controlled manually using a gate valve.
 
 \subsection{Blow-off limit detection}

LBO is determined experimentally by reducing the fuel supply while maintaining a constant air flow rate and making the mixture progressively leaner, until the flame visually blows out. An identical method is used to determine rich blow-off. Fuel supply is increased, while keeping a constant air flow rate, to make the mixture richer, until the flame lifts off and is no longer stabilized by the swirling flow field. For the case of rich blow-off, the flame is observed to continue to burn outside the combustor through entrainment of external air. 

\section{Test Cases}
The test matrix shown below summarizes combustor operating conditions at which measurements were acquired using the schematic shown in Fig.~\ref{fig:TstRg}. All tests were carried out at atmospheric pressure. Measurements were carried out at ambient temperature and at inlet air temperatures of 50$^{\circ}$C and 100$^{\circ}$C. \ce{NH3} volume \% was varied from 0-80\%.
\begin{table}[h]
\begin{tabular}{|c|c|c|c|c|c|}
\hline
$\phi$ & \ce{NH3}  & \ce{CH4} & Air flow & Air preheat & Power output\\ 
\hline
0.7-1.2 & 0-80 & 100-20 & 7  & 20, 50, 100  & 12-22\\ \hline
& Volume \% & Volume \% & SCFM & $^{\circ}$C & kW \\ \hline
\end{tabular}
\caption{Range of variation of combustor operating conditions.}
\end{table}

\section{Reactor network simulations}
Reactor network simulations have been widely used to rapidly evaluate performance characteristics of swirl combustors~\cite{park2013prediction,kaluri2018real,novosselov2006chemical}. The basis of reactor network simulations is to simplify the complex 3-D turbulent combustion process occurring in the swirl combustor, such that the entire process can be represented by a network of idealized combustor modules. The combustor modules simulate the flow physics and chemical kinetics occurring in different physical zones of the combustor. The network permits the modules to communicate with each other providing the ability to exchange mass, species, and energy between the modules. Typically, reactor networks are built using two basic combustor modules, viz., a Perfectly Stirred Reactor (PSR), and a Plug Flow Reactor (PFR). A PSR is homogeneously mixed with chemical reactions taking place uniformly throughout the mixture and is useful in representing combustor zones experiencing turbulent, premixed regions undergoing chemical reaction. It is thus 0D in space and operates at a steady-state condition. A PFR is a 1-dimensional reactor with no longitudinal mixing, with specified inlet composition, and is used to model combustor zones having continuous axial flow.

Reactor networks of varying complexity ranging from 2-3 reactors to configurations involving hundreds to thousands of reactors have been proposed in previous work~\cite{stagni2014fully,yousefian2017review}. Determining the number, type, and volume of reactors (as corresponding to the physical combustor domain) is a key task, which has been carried out in previous work using two preferred techniques. In the first approach, images acquired from experiments such as those showing the flow field using Particle Image Velocimetry (PIV) visualizations are processed to detect the regions of mixing, recirculation, and combustion~\cite{rao2008chemical}. In the second approach, reactive/non-reactive CFD simulations are conducted and simulation results are utilized to estimate the volumes of different flow regions~\cite{stagni2014fully}. Apart from volume, the recirculation fraction of combustion gas is a required input for the reactor network. In the present work, a 3-D non-reacting flow simulation of the combustor geometry with corresponding inlet and operating conditions is used to determine volume and recirculation mass fractions for the reactor network. Details of the CFD flow simulation are discussed next followed by a description of the reactor network. 

\subsection{Flowfield modeling}
 A 3-D, steady, non-reacting flow simulation for the combustor geometry, as used in the experiment, is conducted in Ansys-Fluent. Operating conditions for the simulations include reactant flow rates consistent with experiments and ambient pressure. Flow simulations were conducted for a methane-air mixture with $\phi$=1. The objective of the simulation is to generate the velocity flow field to be used to determine the architecture of the reactor network. As such, only minor differences are expected in results of the non-reacting simulations with changes in fuel composition. In the experiment, gas temperature is expected to vary in the stream wise direction, with lowest temperature expected at the inlet and peak temperature occurring at the reaction zone. The temperature is expected to increase between inlet and reaction zone due to recirculation, while temperature is expected to decrease between the reaction zone and outlet, due to heat loss. To include the effect of streamwise temperature variation, inlet temperature for CFD simulations was set at 1000 K, which can be physically interpreted as an average/ representative temperature of the flow field. Further, computational trials conducted with higher inlet temperatures of 1500 and 2000 K did not show significant variation in the recirculation volume and mass fractions. 
 
 \begin{figure}[H]
	\centering
	\includegraphics[scale=0.45]{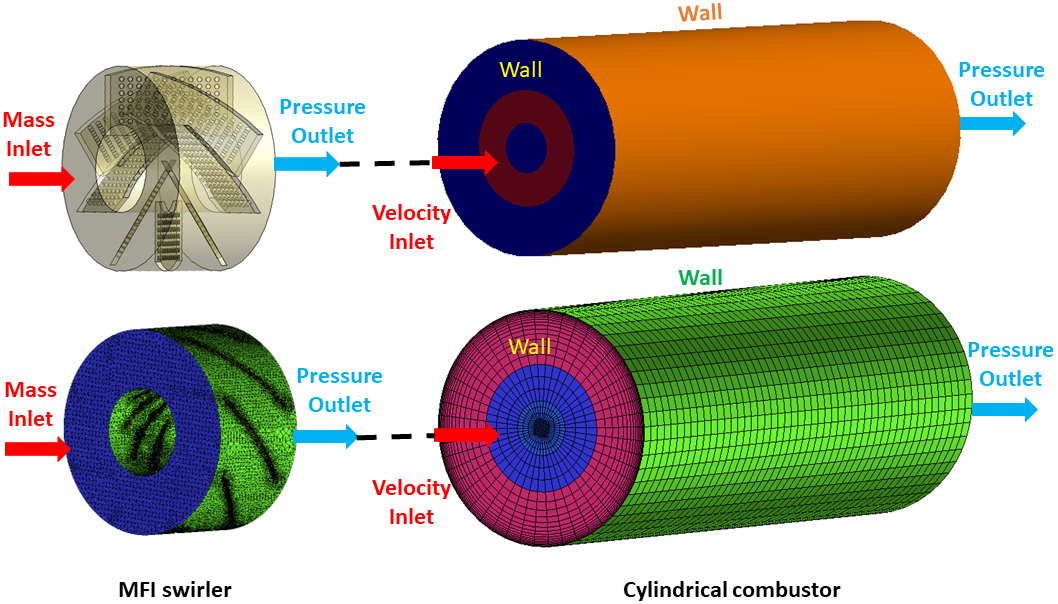}
	\caption{Swirler and combustor geometry and mesh.}
	\label{fig:SwirlerCombustorModel}
\end{figure}
 For the ease of computation, the combustor set up was divided into two parts - inlet MFIS swirler and cylindrical combustion chamber. Figure~\ref{fig:SwirlerCombustorModel} shows the geometry for the swirler module at the combustor inlet. For swirler flow simulation, mass flow rate at the inlet was specified while pressure was prescribed at the outlet. To simulate the fuel injection at the wall of swirler vanes, a volumetric mass source term was added in the computational cell adjacent to the swirler vanes (Wall), where the integrated fuel injection rate  corresponds to the fuel mass flow rate employed in the experiment. Figure~\ref{fig:SwirlerCombustorModel} also shows the geometry of the cylindrical combustion chamber. For the combustor simulation, velocity, species mass fraction, and turbulence field from the outlet of swirler was patched to the combustor inlet and pressure was prescribed at the outlet. Combustor and swirler walls are assumed to be thermally insulated. Details of the numerical model and discretization schemes are provided in Table~\ref{Tab:NumModels}. Due to the complexity of the computational domain, swirl geometry was discretized into unstructured tetrahedral elements, while combustor geometry was divided into structured hexahedral elements. Based on computational trials, to ensure a tractable computational load, the swirler simulations were conducted on a mesh with $\approx$ 4 million elements, with orthogonal quality of $\approx$ 0.4, while a mesh with $\approx$ 12 million orthogonal elements was employed for the combustor flow simulations.

\begin{table}[h]
\caption {Numerical models and discretization.}
\centering
\small
\tabcolsep=0.11cm
\begin{tabular}{| p{6cm} | p{6cm} |} 
\hline
   \textbf{Numerical aspect} &\textbf{Solution method} \\ \hline
  PV coupling &	SIMPLEC\\\hline
  Turbulence & k-$\epsilon$, Standard wall function\\\hline
  Gradient &	Green Gauss Node based\\\hline
  Pressure &	PRESTO!\\\hline
  Momentum, Energy (convective) & Second order upwind\\\hline
\end{tabular}
\label{Tab:NumModels} 
\end{table}

\begin{figure}[H]
	\centering
	\includegraphics[scale=0.5]{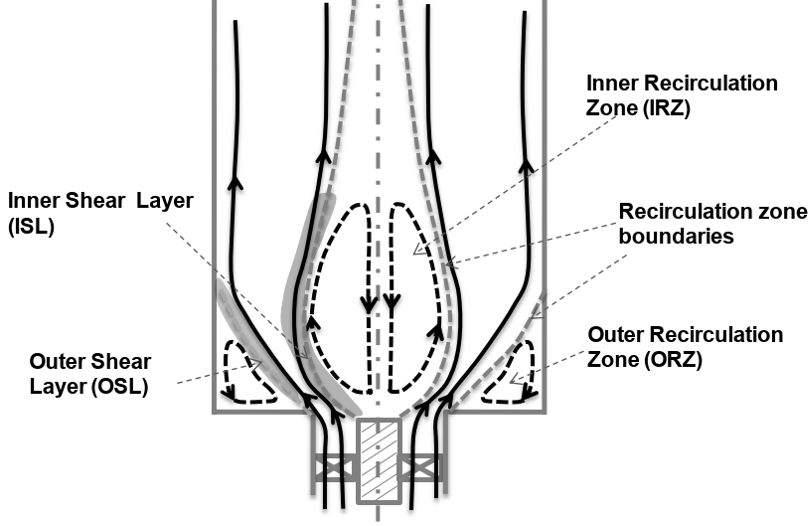}
	\caption{Schematic of a swirl flame \cite{bompelly2013lean}.}
	\label{fig:swirlsche}
\end{figure}

\begin{figure}[H]
	\centering
	\includegraphics[scale=0.35]{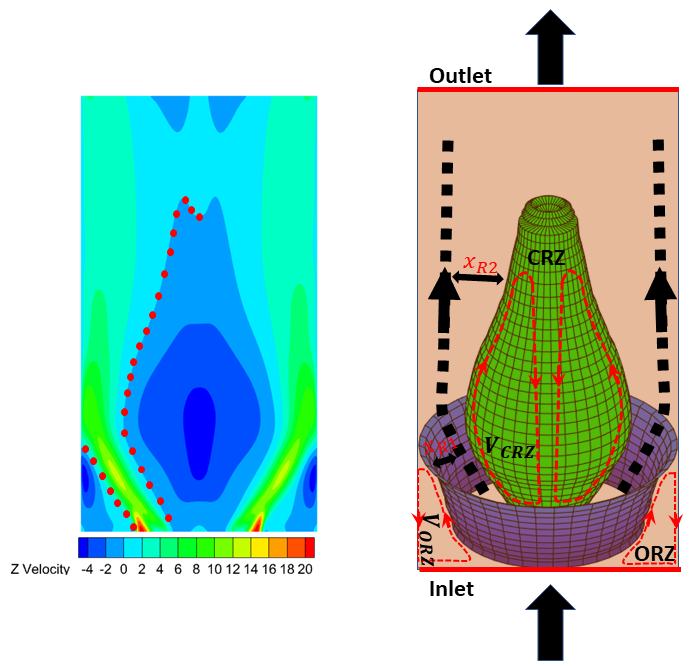}
	\caption{Streamwise velocity contours and mapping of recirculation zone using non-reacting flowfield simulation.}
	\label{fig:SurfaceMapping}
\end{figure}

Based on previous studies, a swirl combustor flow field is expected to have two predominant recirculation zones as shown in Fig. \ref{fig:swirlsche}: Central/Inner Recirculation Zone (CRZ/IRZ) along the axis of the combustor and Outer Recirculation Zone (ORZ) at the lower corner of the combustor.
To evaluate the respective volumes, contours of the streamwise velocity field at mid-plane of the cylindrical combustor were extracted from the simulation results and the interface/boundary points of the recirculation zones (CRZ/IRZ, ORZ) were manually marked, as shown in Figure~\ref{fig:SurfaceMapping}. Assuming axisymmetry, the marked points were revolved about the axis of the combustor, generating a hypothetical surface marking the recirculation zone boundaries and the total volume enclosed by the CRZ and ORZ surfaces was identified as the combined volume of the recirculation zone.       

To quantify the rate of mass exchange between the recirculating and non-recirculating flow regions, uniformly spaced points were marked over the recirculation surface and the velocity components, at these points, were interpolated from the CFD results. Also, with the aforementioned boundary points as vertices, the recirculation boundary surface was discretized into quadrilateral elements, as shown in Fig.~\ref{fig:SurfaceMapping} and for each surface element, the normal mass flow rate was evaluated, using the interpolated velocity field and gas density. The summation of positive or negative mass flow rates over all the surface elements was recorded as an estimate for mass exchange to/from the recirculating flow zones.  

\subsection{Reactor network setup}
\begin{figure}[H]
	\centering
	\includegraphics[scale=0.35]{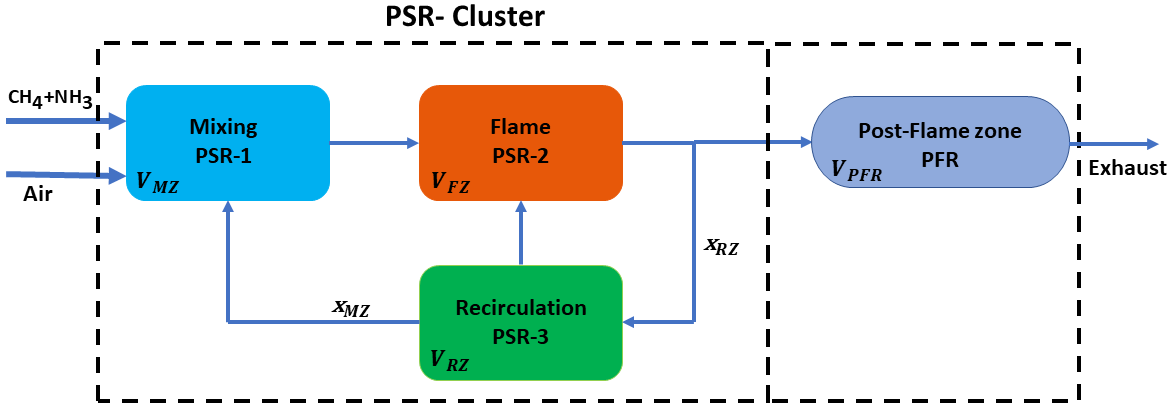}
	\caption{Reactor network simulating the swirl combustor.}
	\label{fig:rnt}
\end{figure}

Figure~\ref{fig:rnt} shows the configuration of the reactor network as derived from the flowfield simulation to simulate the performance of the swirl combustor from the experimental setup. The network, implemented in Ansys Chemkin-Pro, consists of a three PSR reactor cluster, connected to a PFR. 
The Mixing PSR (PSR-1) identifies the inlet MFIS swirler $\&$ initial mixing region of the combustion chamber. The Flame PSR (PSR-2) identifies the combustion zone with positive streamwise velocity. The recirculation PSR (PSR-3) identifies the flow region with recirculation. Volume of each PSR and representative dimensions, including length and diameter of the PFR are obtained from the flowfield simulations described in the previous section. CFD simulation results are also used to determine the fraction of mass recirculated from the Flame and Recirculation PSR's. Volume of reactors and recirculation fractions are listed in Table~\ref{Tab:reacpar}.
\begin{table}[h]
\begin{tabular}{| p{4cm} | p{8cm} |} 
\hline
   \textbf{Input parameter} &\textbf{Set value} \\ \hline
   {} & $V_{MZ}=31$ \\\cline{2-2} 
  Volume [$cm^3$] & $V_{FZ}=630$ \\\cline{2-2} 
  {} & $V_{RZ}=161$ \\\cline{2-2}
  {} & $V_{PFR}=535 \quad (L=10 cm, D=8 cm)$ \\\hline
  Recirculation fraction & $x_{RZ}=0.4, \quad x_{MZ}=0.05$\\\hline
\end{tabular}
\caption{Reactor network parameters.}
\label{Tab:reacpar} 
\end{table}

Additionally, the reactor network requires a chemical reaction mechanism for methane-ammonia mixtures of interest to this work, inlet fuel (\ce{CH4} + \ce{NH3}) and 
air mass flow rates, inlet reactant temperature, and rates of heat loss from the PSR and PFR.
To simulate the combustion reaction, the reaction mechanism developed in the work of Okafor~\cite{okafor2019measurement} is employed, which was an adaptation of GRI 3.0~\cite{smith1995gri} and Tian's~\cite{tian2009experimental} chemical mechanisms with 59 species and 356 reactions. In previous numerical studies, Okafor's reaction mechanism has been shown to provide a good estimation for production of \ce{NOx} and other combustion products in comparison to corresponding experiments. Based on current experiments, simulations were conducted for an inlet mass flow rate of 4.3 gm/s, for three inlet temperatures - $20$, $50$ and $100^0$C.
 During trial simulations using the reactor network, it was observed that blow-off and emission predictions are sensitive to heat loss in each reactor. Heat loss is the only free variable in the reactor network simulations, given that all other parameters are constrained by the experimental conditions and PSR/PFR attributes as extracted from the non-reacting flow simulation. In previous work using reactor networks~\cite{strakey2007investigation,haynes2008trapped}, simulation settings are tuned by adjusting available free parameters to anchor the network settings and align simulation results to experimental data for a chosen data point. Heat loss is used in a similar manner in this work. This approach is supported by the main objective of reactor network simulations in this work, which is to conduct sensitivity analyses and study reaction pathways responsible for \ce{NOx} generation.
 
 
Equivalence ratios for rich and lean blow-off were numerically evaluated for the reactor network presented in Fig.\ref{fig:rnt}. The procedure starts with generation of a stable flame using a starting $\phi$ of 0.7 for LBO and 1 for RBO detection. Next, by adding continuations in Chemkin-Pro, $\phi$ is decreased for LBO or increased for RBO in steps of 0.01. The use of continuations implies that each successive solution uses the solution for the previous $\phi$ as the initial condition.  

\section{Results and Discussion}
\subsection{Lean blow-off studies}
\begin{figure}[H]
	\centering
	\includegraphics[scale=0.5]{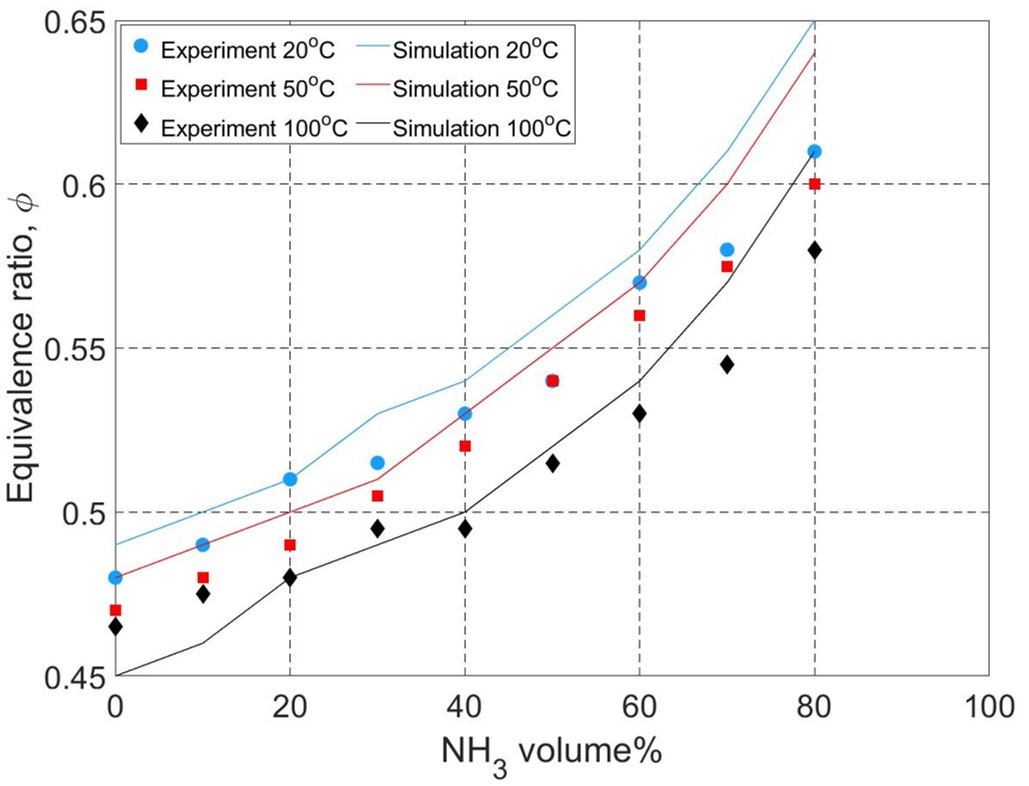}
	\caption{Equivalence ratio for lean blow out (LBO) as a function of ammonia addition.}
	\label{fig:LBO}
\end{figure}

\begin{figure}[H]
	\centering
	\includegraphics[scale=0.65]{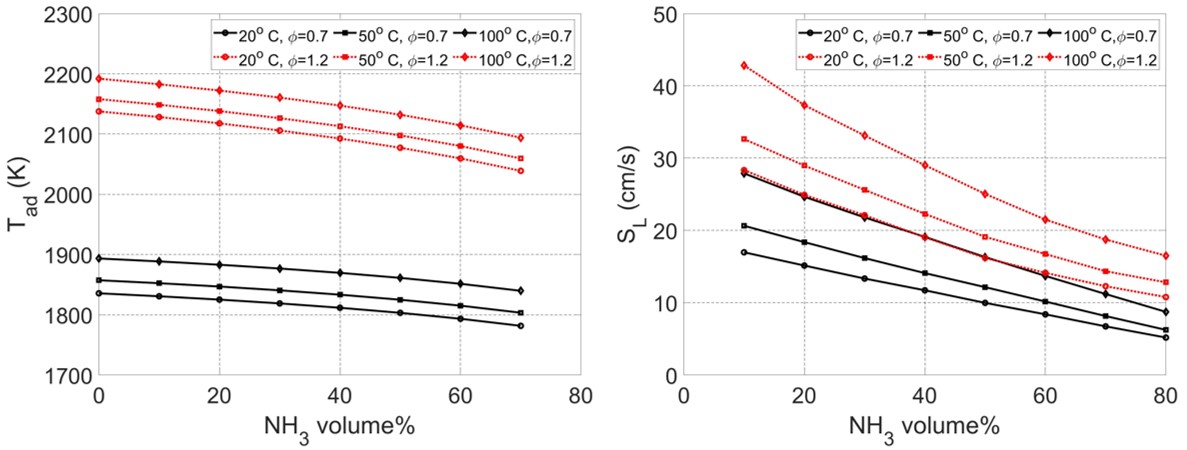}
	\caption{Adiabatic flame temperature and laminar flame speed.}
	\label{fig:Tad_Sl}
\end{figure}

\begin{figure}[H]
	\centering
	\includegraphics[scale=0.65]{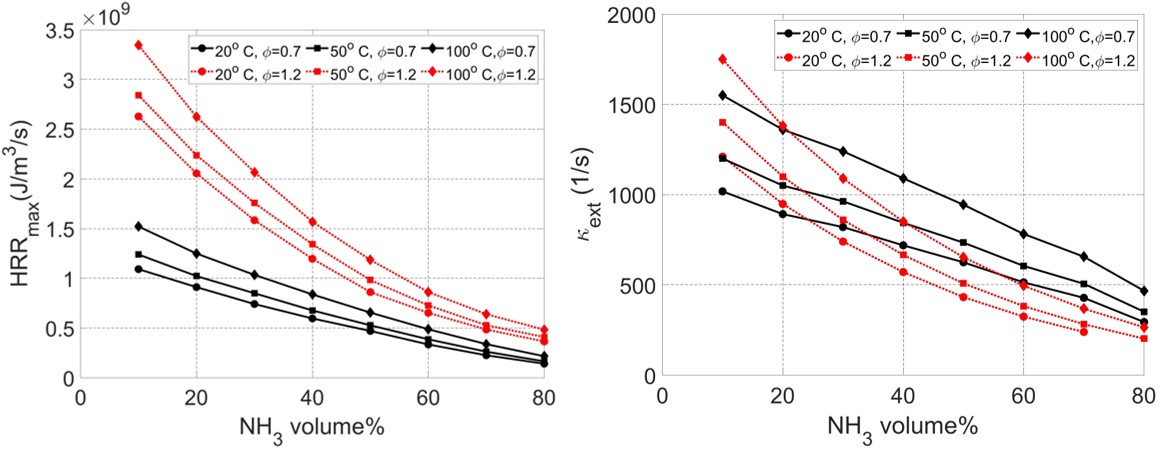}
	\caption{Maximum heat release rate and extinction strain rate.}
	\label{fig:HRR_Kext}
\end{figure}

Figure~\ref{fig:LBO} shows $\phi$ for LBO as a function of \ce{NH3} volume \% for three inlet air temperatures of 20, 50, and 100$^0$C. Results are presented from experiments and reactor network simulations. Lowest $\phi_{LBO}$ for any \ce{NH3} volume \% is found to occur for 100\% \ce{CH4} (0\% \ce{NH3}) and increases with addition of ammonia. There are several factors contributing to decrease in flame stability with increasing ammonia addition.
 As shown in Fig.~\ref{fig:Tad_Sl}, adiabatic flame temperature ($T_{ad}$) and laminar flame speed ($S_l$) decrease with ammonia addition. This is primarily driven by the lower energy content of ammonia (22.5 MJ/kg) as compared to methane (55.5 MJ/kg). This results in a lower heat release rate, as shown in Fig.~\ref{fig:HRR_Kext}, which has a direct influence on the stabilization mechanism for swirl flames. As discussed by Stohr~\cite{stohr2011dynamics}, two particular flame zones, a helical flame and the flame root play a key role in the dynamic blow-off process occurring in swirl flames. The flame root anchors the flame close to the swirler exit and is identified as the point where fresh reactants come in contact with burned gases from recirculation zones. The recirculation zones providing a source of heat and radicals to the fresh gas form a key stabilization mechanism for swirl flames. The flame zone is a region of high strain and studies by Stohr~\cite{stohr2011dynamics}, Ji~\cite{ji2022experimental}, and Bohm~\cite{bohm2009time} have noted the extinction at the flame root to be responsible for ensuing blow-off. With addition of ammonia, and corresponding decrease in heat release rate, energy content of the recirculating burned gas decreases, making the flame root susceptible to extinction. Extinction strain rates ($\kappa_{ext}$) plotted in Fig.~\ref{fig:HRR_Kext} show that addition of ammonia significantly reduces $\kappa_{ext}$. These factors combined with the high strain rate at the flame root leads to local extinction and blow-off. Increase in inlet air temperature is seen to improve flame stability through a corresponding increase in $T_{ad}$, $S_l$, and $\kappa_{ext}$, as observed in Fig.~\ref{fig:Tad_Sl} and Fig.~\ref{fig:HRR_Kext}. Reactor network simulation predictions, wherein a single value of heat loss is used for all cases, are observed to have a good agreement with measurements for $\phi_{LBO}$.

\subsection{Rich blow-off studies}
\begin{figure}[H]
	\centering
	\includegraphics[scale=0.4]{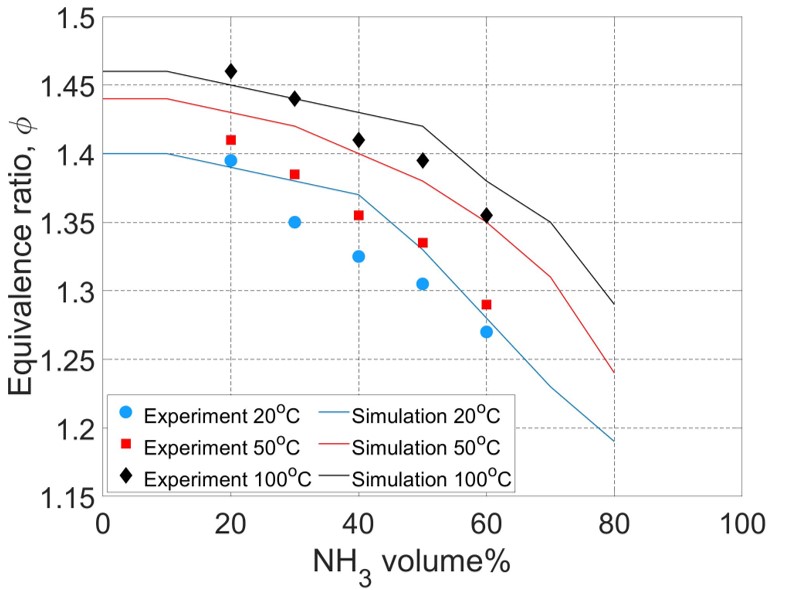}
	\caption{Equivalence ratio for rich blow out (RBO) as a function of ammonia addition.}
	\label{fig:RBO}
\end{figure}

Figure~\ref{fig:RBO} shows $\phi_{RBO}$ obtained from measurements and predicted by reactor network simulations. Results are plotted as a function of \ce{NH3} volume \% for different values of inlet air temperature. Similar to results for $\phi_{LBO}$, highest flame stability corresponding to maximum value of $\phi_{RBO}$ is found for pure methane flames. With increase in ammonia content, $\phi_{RBO}$ is found to decrease due to similar reasons as discussed earlier for $\phi_{LBO}$. Decrease in $T_{ad}$, $S_l$, and $\kappa_{ext}$, contribute to decrease in flame stability and are responsible for driving local extinction at the flame root leading to blow-off.
 Results presented in Fig.~\ref{fig:RBO} show that the combustor could be operated on fuel mixtures with \ce{NH3} volume \% down to 20 \% (80\% \ce{CH4}) with a bulk Re of $\sim$ 4600 prior to rich blow-off. This represents a significant increase in the operational limits for methane-ammonia-air swirl flames. In prior work by Khateeb~\cite{khateeb2020stability}, the occurrence of flashback limited swirl combustor operation to 40 \% \ce{NH3} (60\% \ce{CH4}) with a bulk inlet Re of 7000. For lower \ce{NH3} content, flame speed exceeds jet velocity resulting in flashback. The MFIS however, generates fuel-air mixing at the swirler vanes, thus rendering the issue of flashback to be almost non-existent. Further, water cooling provided to the swirler ensures that flame speed would be significantly reduced at the swirler due to heat loss, thus ensuring no flashback induced thermal degradation of the swirler module. An increase in inlet air temperature is seen to further increase the operational limits of the combustor leading to higher $\phi_{RBO}$. Once again, good agreement is observed between experimental measurements and reactor network simulation predictions for $\phi_{RBO}$.

\subsection{Stability limits}
\begin{figure}[H]
	\centering
	\includegraphics[scale=0.5]{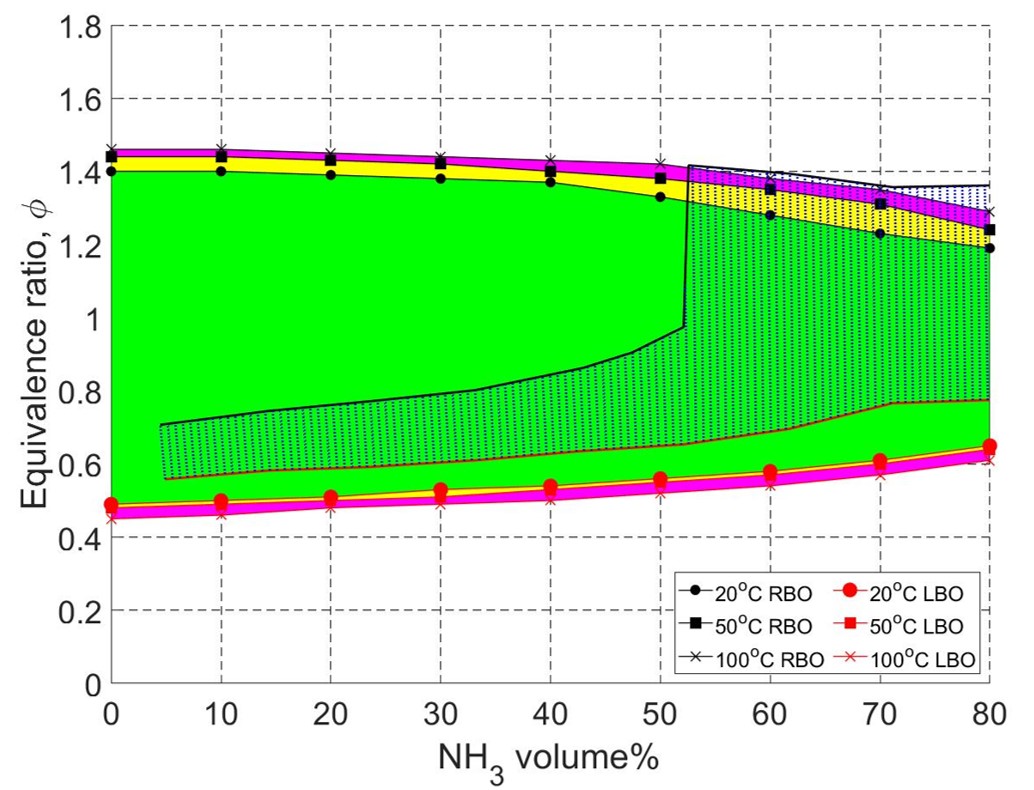}
	\caption{Stability limits for combustor operation on methane-ammonia-air mixtures based on rich and lean blow-off limits. Dotted region corresponds to results by Khateeb~\cite{khateeb2020stability}.}
	\label{fig:Stability}
\end{figure}

Figure~\ref{fig:Stability} shows LBO and RBO limits as determined through reactor network simulations overlaid to illustrate the operational limits for the combustor utilizing the MFIS module. Also overlaid on the plot is the stability limit established by Khateeb~\cite{khateeb2020stability} using a conventional swirler operating at ambient inlet air temperature. Effects of air preheating are also included in the results presented in Fig.~\ref{fig:Stability}. As can be observed, use of the MFIS module allow for significant expansion of stability limits over a conventional swirler. This is particularly true for operation at rich mixtures with high methane content, which is susceptible to flashback issues for a conventional swirler. 
Further, LBO limits are observed to be reduced for the entire range of \ce{NH3} volume \% through the use of the MFIS module. Finally, air preheating is found to achieve slight increases in stability limits at lean and rich conditions, but this strategy has to be considered in tandem with its effect on \ce{NOx} production, which will be discussed next.

\subsection{NO emissions}
\begin{figure}[H]
	\centering
	\includegraphics[scale=0.6]{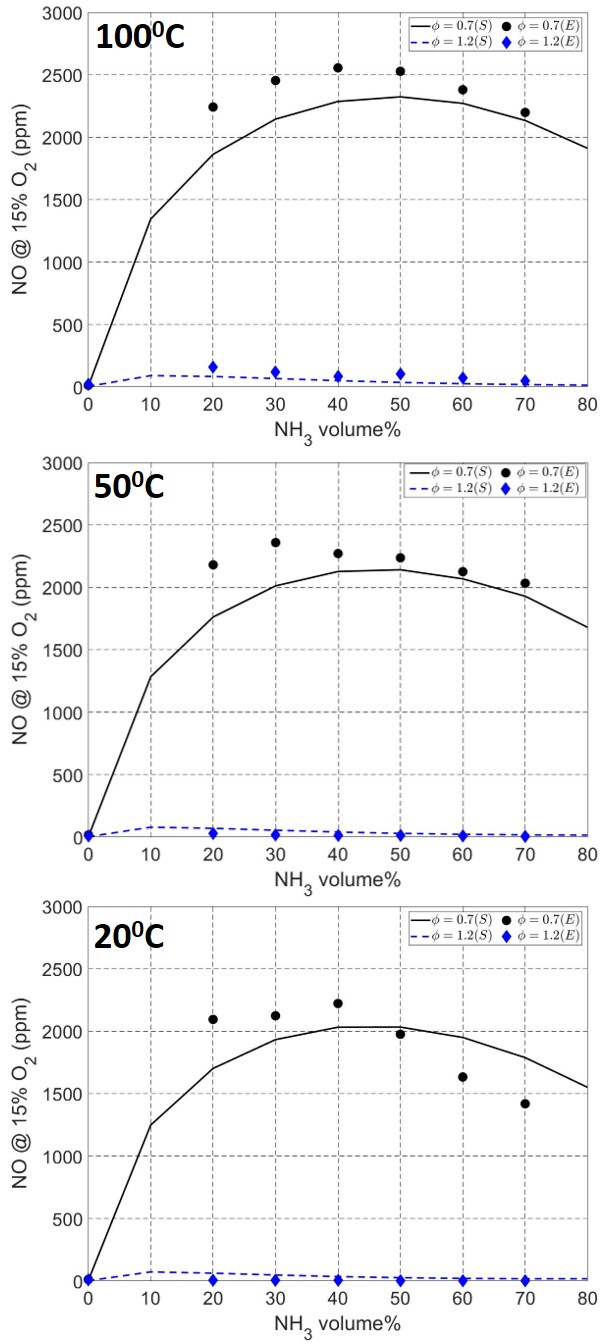}
	\caption{NO emissions from experiments (\textit{E}) and reactor network simulations (\textit{S}).}
	\label{fig:NOx}
\end{figure}

Figure~\ref{fig:NOx} shows \ce{NO} emissions as a function of \ce{NH3} volume \% for lean ($\phi$=0.7) and rich ($\phi$=1.2) mixtures. Results are presented for the three inlet air temperatures combining experimental measurements using the exhaust gas analyzer and predictions from reactor network simulations. Previous work has shown that \ce{NO2} emissions account for a small amount of total \ce{NOx} (\ce{NO}+\ce{NO2}) emissions. This was also confirmed by reactor network simulation results in the present work, showing \ce{NO} to be less than $\sim$2\% of total \ce{NOx}. Thus, only \ce{NO} emissions are reported in this work. \ce{NO} measurements are corrected to 15\% \ce{O2} content and are reported on a dry basis in parts per million (ppm) units.
 A significant drop is noted in \ce{NO} concentration going from the lean ($\phi$=0.7) to rich ($\phi$=1.2) mixture for all temperatures. For both mixture conditions, \ce{NO} emissions increase with ammonia addition, peaks at around 40\%, and then decreases with increasing ammonia addition. With increase in inlet air temperature, \ce{NO} generation is seen to increase for both mixture conditions. Trends for \ce{NO} concentration as observed in Fig.~\ref{fig:NOx} are now analyzed using the contributions to \ce{NO} production from different reaction pathways. This is facilitated by results obtained from the reactor network simulations.


\begin{table}[h]
\footnotesize
\centering
\begin{tabular}{| p{4.5cm} | p{4.5cm} |} 
\hline
   \textbf{\ce{NO} production pathway} &\textbf{Reactions} \\ \hline
   \multirow{3}{*}{Thermal \ce{NOx}}  & \ce{N + OH $\rightleftharpoons$ NO + H} \\\cline{2-2} 
   {}  & \ce{N + O2 $\rightleftharpoons$ NO + O} \\\cline{2-2} 
   {}  & \ce{N2 + O $\rightleftharpoons$ N + NO} \\\hline
    \multirow{2}{*}{\ce{N2O} pathway}  & \ce{N2O + O $\rightleftharpoons$ 2NO} \\\cline{2-2} 
    {}  & \ce{N2 + H $\rightleftharpoons$ NH + NO} \\\hline
      \multirow{3}{*}{\ce{HNO} pathway}  & \ce{HNO ( + M) $\rightleftharpoons$ NO + H ( + M)} \\\cline{2-2} 
   {}  & \ce{HNO + OH $\rightleftharpoons$ NO + H2O} \\\cline{2-2} 
   {}  & \ce{HNO + O2 $\rightleftharpoons$ NO + HO2} \\\hline 
      \multirow{5}{*}{\ce{NHi} pathway}  & \ce{NH + O$\rightleftharpoons$ NO + H} \\\cline{2-2} 
   {}  & \ce{NH + O2 $\rightleftharpoons$ NO + OH} \\\cline{2-2} 
   {}  & \ce{NH + NO $\rightleftharpoons$ N2 + OH} \\\cline{2-2}
   {}  & \ce{NH2 + NO $\rightleftharpoons$ NNH + OH} \\\cline{2-2}
   {}  & \ce{NH2 + NO $\rightleftharpoons$ N2 + H2O} \\\hline
   \hline
\end{tabular}
\caption{\ce{NO} production pathways and corresponding reactions.}
\label{Tab:NOpathways} 
\end{table}
Four main reaction pathways for \ce{NO} formation in methane-ammonia-air combustion have been identified as: Thermal \ce{NOx}, \ce{N2O} pathway, \ce{HNO} pathway, and \ce{NHi} pathway~\cite{li2019analysis}. Table~\ref{Tab:NOpathways} summarizes the key chemical reactions involved in each of the four NO production pathways.

 \begin{figure}[H]
 	\centering
 	\includegraphics[scale=0.75]{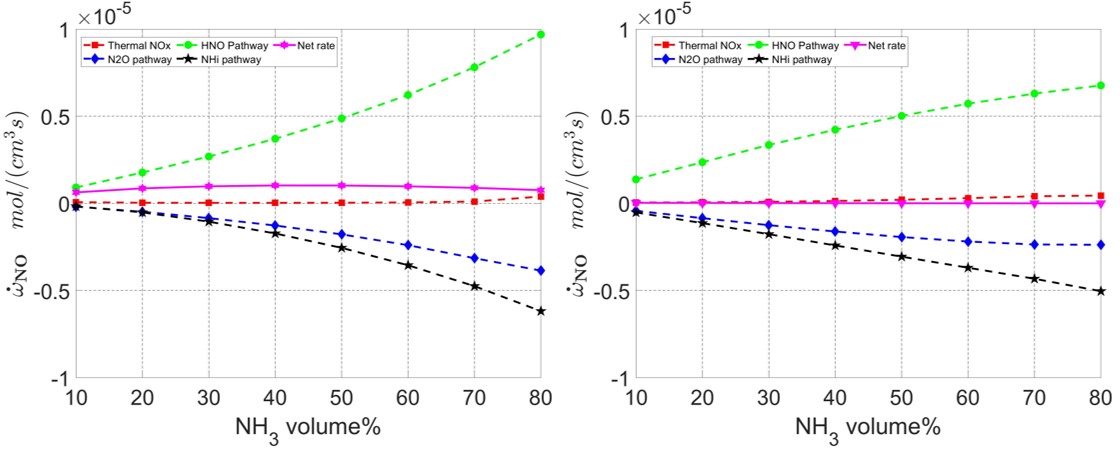}
 	\caption{NO production rates through the different reaction pathways as listed in Table~\ref{Tab:NOpathways} at lean ($\phi$=0.7, \textit{Left}) and rich ($\phi$=1.2, \textit{Right}) conditions.}
 	\label{fig:PathAnalysis}
 \end{figure}
Figure~\ref{fig:PathAnalysis} shows the contribution to \ce{NO} production from the four different pathways listed in Table~\ref{Tab:NOpathways} as well as the net production rate. \ce{NO} production rate contributions from the different pathways are obtained as the sum total from the PSR cluster. 
Results in Fig.~\ref{fig:PathAnalysis} are presented as a function of \ce{NH3} volume \% for lean ($\phi$=0.7) and rich ($\phi$=1.2) conditions. \ce{NO} production at lean mixtures is seen to be dominated by \ce{HNO} pathway, while \ce{NO} consumption is influenced by \ce{N2O} and \ce{NHi} pathways. The thermal \ce{NO} pathway (Zel'dovich mechanism) is seen to not play a significant role for either case shown in Fig.~\ref{fig:PathAnalysis}. The OH radical plays a key role in \ce{NO}  production primarily through the \ce{HNO + OH $\rightleftharpoons$ NO + H2O} reaction. At the same time, the OH radical is also key to \ce{NH2} and \ce{NH} production through hydrogen atom abstraction reactions (\ce{NH3 + OH and \rightleftharpoons NH2 + H2O} and \ce{NH2 + OH and \rightleftharpoons NH + H2O}). OH production rate is seen to stay almost constant with increasing \ce{NH3} volume \%. However, increasing \ce{NH3} volume \% causes increasing competition for OH radical between the NO producing HNO reaction and the hydrogen abstraction reactions. The net result is seen to be a slower rate of increase of NO production from the HNO pathway and an increased rate of consumption of NO through \ce{N2O} and \ce{NHi} pathways. These factors lead to the overall decrease in \ce{NO} concentration with increasing \ce{NH3} volume \% as observed in the results presented in Fig.~\ref{fig:NOx}. 
With increase in inlet air temperature, reaction rates increase consistently for all \ce{NO} production pathways, resulting in an increase in \ce{NO} concentration. At rich mixture conditions, \ce{NO} production continues to be dominated by the \ce{HNO} pathway. However, significant availability of \ce{NH3} at the rich conditions promotes the formation of \ce{NH2} and NH through hydrogen atom abstraction~\cite{miller1983kinetic}. Presence of large amounts of NH and \ce{NH2} promotes \ce{NO} consumption through \ce{NHi} pathways resulting in significantly lower \ce{NO} concentration at rich conditions as observed in Fig.~\ref{fig:NOx}.


\section{Conclusion}
The consideration of ammonia, a carbon-free fuel, has been spurred by efforts to reduce greenhouse gas emissions and minimize the impact of hydrocarbon fuel combustion-based power generation on global warming and climate change. In previous studies addressing premixed swirl combustors, methane addition was suggested to improve ammonia combustion characteristics. While addition of methane improves flame stability, it is found to increase \ce{NOx} emissions. A staged combustion approach with a rich primary and lean secondary stage has been recommended to reduce \ce{NOx} emissions while ensuring complete fuel oxidation. Success of these combustion approaches for methane-ammonia-air mixtures relies on the ability of a combustor to operate with wide stability limits permitting the use of fuel-rich and fuel-lean conditions as needed to achieve optimal performance. The present work evaluates two strategies to achieve increase in operational limits and their corresponding influence on \ce{NOx} emissions. The first strategy involves the use of multi-point fuel injection by employing a novel swirler module to achieve increased homogeneity in fuel-air mixing, while alleviating flashback concerns. The second strategy involves preheating of air at the combustor inlet. A model combustor setup equipped with the novel swirler module and air preheating is used to study combustion of premixed methane-ammonia-air flames. A chemical reactor network is developed using non-reacting simulations of the combustor flow field to obtain insight into the reaction pathways responsible for \ce{NO} generation. Studies considering flame blow-off limits and corresponding \ce{NO} emissions were conducted at atmospheric pressure with three different inlet air temperatures (20$^0$C, 50$^0$C, 100$^0$C) for fuel lean ($\phi$=0.7) and rich ($\phi$=1.2) conditions. Measurement and simulation results are analyzed by considering flame heat release rate, extinction strain rate, laminar flame speed, and adiabatic flame temperature. Reaction pathways leading to \ce{NO} production are analyzed to explain \ce{NO} emissions as a function of  equivalence ratio and percentage ammonia addition. The important findings of this study are as follows:
\begin{enumerate}

    \item Equivalence ratio for lean blow-off is seen to increase with ammonia addition to the fuel mixture. Given its lower energy content, ammonia addition leads to decrease in heat release rate and energy content in the recirculating burned gases causing the flame to be increasingly susceptible to blow-off. Increase in air inlet temperature improves lean blow-off through increase in adiabatic flame temperature, flame speed, and extinction strain rate as shown by experimental measurements consistent with reactor network simulations.
	
	\item Equivalence ratio for rich blow-off decreases with increase in ammonia addition. Increase in inlet air temperature is seen to delay rich blow-off in a manner similar to lean blow-off and is attributed to the same factors, i.e., increase in adiabatic flame temperature, flame speed, and extinction strain rate. 
	
	\item Improved fuel-air mixing, reduced mixing length ($\sim$0.5"), and cooling of the swirler module renders the combustor to be highly resistant to flashback. These advantages due to the novel swirler module along with use of preheated air permit significant enhancement of the operability limits of the combustor. These advantages can be leveraged for an efficient design of a two-stage combustor.
	
	\item \ce{NO} emissions are very high at lean conditions, exceeding 2000 ppm. A significant reduction in \ce{NO} emissions (under 200 ppm) is seen to be achieved by rich operation. Reaction path analysis reveals the \ce{HNO} pathway to be key for \ce{NO} production at lean conditions. For rich mixtures, consumption of \ce{NO} through the \ce{NHi} pathway, promoted by availability of NH and \ce{NH2} through hydrogen abstraction of excess \ce{NH3} is found to be responsible for lower \ce{NO} concentration.  For rich and lean mixtures, increase in inlet air temperature is found to increase \ce{NO} concentration primarily through increase in reaction rates. \ce{NO} emissions for lean conditions at all temperatures is found to increase with ammonia addition of up to $40\%$ and then decrease with further ammonia addition. This is found to be caused by increased competition for OH radicals leading to higher NO consumption by the \ce{NHi} pathway and lower \ce{NO} production through the \ce{HNO} pathway. The thermal \ce{NO} pathway is observed to have no significant influence on \ce{NO} formation at lean and rich mixture conditions.
	\item Increase in combustor stability limits through inlet air preheating is not found to be significant. Additionally, higher inlet air temperatures are seen to increase NO emissions, implying that this is likely not a viable strategy for methane-ammonia-air combustion.
	\item The use of a three reactor network designed using flow field information generated by a non-reacting flow simulation, as well as heat loss as a tuning parameter to anchor the reactor network simulations, is found to give good agreement with experimental measurements.
\end{enumerate}

\section{Acknowledgements}
This work was funded in part by the LaSPACE Research Award Program (RAP) and the authors gratefully
acknowledge support of this work by NASA EPSCoR.

\bibliographystyle{unsrtnat}
\bibliography{biblio1}




\end{document}